\RequirePackage{ifpdf}
\ifpdf 
\documentclass[pdftex]{sigma}
\else
\documentclass{sigma}
\fi

\numberwithin{equation}{section}

\newtheorem{Theorem}{Theorem}[section]
\newtheorem{Corollary}[Theorem]{Corollary}
\newtheorem{Lemma}[Theorem]{Lemma}
\newtheorem{Question}[Theorem]{Question}
{\theoremstyle{definition}
\newtheorem{Remark}[Theorem]{Remark}
\newtheorem{Definition}[Theorem]{Definition}
}

\def\1{\mathrel{\mathbf{1}}}

\newcommand{\M}{\mathbf{M}}
\newcommand{\bA}{\mathbf{A}} 

\newcommand{\mG}{\mathcal{G}} 

\newcommand{\act}{\curvearrowright}

\newcommand{\PG}{\ensuremath{\mathbf{PG}}}

\newcommand{\PSL}{\ensuremath{\mathbf{PSL}}}

\begin{document}

\allowdisplaybreaks

\renewcommand{\PaperNumber}{080}

\FirstPageHeading

\ShortArticleName{The 2-Transitive Transplantable Isospectral Drums}

\ArticleName{The 2-Transitive Transplantable Isospectral Drums}

\Author{Jeroen SCHILLEWAERT~$^\dag$ and Koen  THAS~$^\ddag$}

\AuthorNameForHeading{J.~Schillewaert and K.~Thas}

\Address{$^\dag$~Department of Mathematics, Free University of Brussels (ULB),\\
\hphantom{$^\dag$}~CP 216, Boulevard du Triomphe, B-1050 Brussels, Belgium}
\EmailD{\href{mailto:jschille@ulb.ac.be}{jschille@ulb.ac.be}}

\Address{$^\ddag$~Department of Mathematics, Ghent University, Krijgslaan 281, S25, B-9000 Ghent, Belgium}
\EmailD{\href{mailto:kthas@cage.UGent.be}{kthas@cage.UGent.be}}
\URLaddressD{\url{http://cage.UGent.be/~kthas/}}

\ArticleDates{Received December 14, 2010, in f\/inal form August 08, 2011;  Published online August 18, 2011}

\Abstract{For Riemannian manifolds there are several examples which are isospectral but not isometric, see e.g.~J.~Milnor [{\em Proc. Nat. Acad. Sci. USA} {\bf 51} (1964), 542]; in the present paper, we investigate pairs of domains in~$\mathbb{R}^2$ which are isospectral but not congruent.
All known such counter examples to M.~Kac's famous question can  be constructed by a~certain tiling method (``transplantability'') using special linear operator groups which act $2$-transitively on certain associated modules.
In this paper we prove that if {\em any}  operator group acts $2$-transitively on the associated module,
no new counter examples  can occur.
In fact, the main result is a corollary of a result on Schreier coset graphs of $2$-transitive groups.}


\Classification{20D06; 35J10; 35P05; 37J10; 58J53}

\section {Introduction}

A very good exposition of the importance of the problem we want to consider in this paper and its history can be found in~\cite{Conw}, see also~\cite{GT}~-- we describe it here in a nutshell. Four decades ago, M.~Kac posed the following famous question in a paper published in Amer. Math. Monthly~\cite{Kac}: ``Can one hear the shape of a drum?''.
Clearly, once you know the shape of a~drumhead, there is a well-established mathematical theory telling you at which frequencies this drumhead can vibrate. The question of Kac is the inverse problem~-- in other words, is it possible to determine the shape of the drumhead from the sound it makes, so {\em from the frequencies at which it vibrates}? It is the popularization of the question as to whether there can exist two non-congruent isospectral domains in the real plane. So what can be inferred on a domain $D$ if one only knows the eigenvalues of the Dirichlet problem for the Laplacian?
Almost immediately, J.~Milnor in \cite{Milnor} produced a pair of 16-dimensional tori that have the same eigenvalues, but dif\/ferent shapes, providing thus a counter example to M.~Kac's question in higher dimension. Other constructions of counter examples can be found in, e.g.~\cite{Bus,BroTse}~-- see \cite{GT}.
However, it took almost 30 years before a counter example in the plane was found; in 1992 C.~Gordon, D.~Webb and S.~Wolpert \cite{GWW} constructed a~pair of regions in the plane that have dif\/ferent shapes but identical eigenvalues. Their construction boils down to a sort of tiling method (see below), and using a technique called ``transplantation'' (see further) it is not so hard to check that the domains indeed have the same eigenvalues for the Dirichlet problem. All known counter examples in the plane are constructed in a similar way~-- see \cite{GT}. Quite amazingly, by a method which dates back to Sunada~\cite{Sunada}, f\/inite groups come into play, and it turns out that all known counter examples arise from special linear groups.

In this paper we take a substantial step towards our goal to show that one cannot expect much more from this tiling method~-- in other words, that (up to possibly a small number of exceptions) all counter examples constructed by this tiling method are known. In particular, we  study involutions in 2-transitive groups, which are known by the classif\/ication of f\/inite simple groups, and the modules on which they naturally act. We obtain a formula concerning f\/ixed points for the groups obtained, indicating they should possess a large number of f\/ixed points. The few cases not eliminated by this strategy are dealt with separately in the proof of Corollary~\ref{MR}.

The $2$-transitivity assumption is not merely based on the fact that all known counter examples (as of August 5th~2011) using this tiling method are ${\mathbf{PSL}}$, since there also are no other examples for small $n$, by a computer search of Okada and Shudo \cite[p.~5921]{OS}. So for $n\leq13$ our conjecture is true. Possibly this has already been extended in view of the increased computer power.

As this computer evidence suggests that all the known counter examples using this tiling method are $\mathbf{PSL}$ for small values of~$n$ and~$q$, we try to prove this in several steps.
The step that $\mathbf{PSL}$ leads to the known values of $n$ and $q$ has already been taken. This paper forms the second step, namely ``$2$-transitive implies~$\mathbf{PSL}$''. The last (and most dif\/f\/icult) step is to prove $2$-transitivity\footnote{One could try to achieve this considering a minimal counter example. Often in group theory, minimal counter examples have more special properties. So the strategy would be to show that if there is an example there is also a 2-transitive one, which would lead us back into $\mathbf{PSL}$. We refer to \cite{KTI} for more on that matter.}.

For background information on permutation groups, in particular 2-transitive groups, we refer to \cite{DM}. Let us just mention that a {\em $2$-transitive} group is a group $G$ acting on a set $X$, such that for any two pairs $(x,y)$, $x\neq y$, and $(x',y')$, $x'\neq y'$, in $X\times X$, there exists an element $g\in G$ such that $x^{g}=x'$ and $y^{g}=y'$.

(For notions not yet introduced, we refer to Section~\ref{dim2}.)

\begin{Theorem}[main result]
\label{main}
Let $D_1$ and $D_2$ be non-isometric isospectral simply connected domains in $\mathbb{R}^2$ which are transplantable.
If the associated operator group is $2$-transitive in its action defined by transplantability,
then it is isomorphic to $\PSL_n(q)$ with $(n,q) \in \{(3,2),(3,3),(4,2),$ $(3,4)\}$, and
$(D_1,D_2)$ belongs to the list mentioned in {\rm \cite[p.~5921]{OS}}.
\end{Theorem}

\section{The Kac problem}
\label{dim2}

\begin{Definition}
To any compact Riemannian manifold $(\M,g)$ one can associate a second-order dif\/ferential operator, the {\em Laplace operator} $\Delta$, def\/ined by
\[
\Delta(f) = - \mathrm{div}(\mathrm{grad}(f))
\]
for $f\in L^{2}(\M,g)$. Two manifolds are {\em isospectral} if they have the same eigenvalue spectrum
(including multiplicities) for the Laplace operator.
\end{Definition}

\begin{Definition}
We say a domain $D \subseteq \mathbb{R}^2$ is {\em simply connected} if it is as a topological space, using the usual metric in $\mathbb{R}^2$. Note that a simply connected
domain is connected (by def\/inition).
\end{Definition}

A celebrated inverse problem posed by M.~Kac \cite{Kac} asks whether simply connected domains in $\mathbb{R}^2$ for which
the sets $\{ \lambda_n \parallel n \in \mathbb{N}\}$ of solutions (eigenvalues) of the stationary Schr\"{o}dinger equation
\[ (\Delta + \Lambda)\Psi = 0 \qquad  \mbox{with}\quad  \Psi\big|_{\text{boundary}} = 0  \]
coincide, are necessarily congruent.
Counter examples were constructed to the analogous question on Riemannian manifolds~-- see \cite{GT}, but for Euclidean domains
the question appears to be much harder.

In 1992 C.~Gordon, D.~Webb and S.~Wolpert \cite{GWW} constructed a pair of simply connected  non-isometric Euclidean
iso\-spectral domains~-- also called ``planar iso\-spectral pairs'' or ``isospectral billiards'' or ``isospectral drums'' (etc.) in the literature.

\subsection{Tiling}

Up to present, all known planar counter examples were constructed by a certain tiling method. Call
such examples isospectral {\em Euclidean TI-domains}.
Up to homothety, only a f\/inite number of examples arises if one f\/ixes the congruence class of the base tile~-- see \cite{GT} for details.

\textbf{Tiling.}
All known isospectral billiards can be obtained by unfolding polygonal-shaped tiles. The way the tiles are unfolded can be specif\/ied by $r$ permutation $N\times N$-matrices $M^{(\mu)}$, $1 \leq \mu \leq r$ and $N \in \mathbb{N}$, associated with the $r$ sides of the $r$-gon ($r \geq 3$):
\begin{itemize}\itemsep=0pt
\item
$M^{(\mu)}_{ij}  =  1$ if tiles $i$ and $j$ are glued by their side $\mu$;
\item
$M^{(\mu)}_{ii}  =  1$ if the side $\mu$ of tile $i$ is on the boundary of the billiard, and
\item
$0$ otherwise.
\end{itemize}
(The number of tiles is $N$.)

One can sum up the action of the $M^{(\mu)}$ in a graph with colored edges:
each copy of the base tile is associated with a vertex, and vertices $i$ and $j$, $i \ne j$, are joined by an edge of color $\mu$ if and only if $M^{(\mu)}_{ij}  =  1$.
In the same way, in the second member of the pair, the tiles are unfolded according to permutation matrices $N^{(\mu)}$, $1 \leq \mu \leq r$.
Call such a colored graph an {\em involution graph} for now.

\begin{Remark}\label{remtree} \qquad
\begin{itemize}\itemsep=0pt
\item[$(i)$]
The consideration of the colored graph as such def\/ined was f\/irst made by Y.~Okada and A.~Shudo \cite{OS}.
We will encounter this graph in a dif\/ferent form at the end of Section~\ref{dim2}.
\item[$(ii)$]
Note that the involution graph is connected, since we consider simply connected domains. Also, if we want the base tile to be of ``any'' shape, there should be no closed circuit in the graph, following an idea of~\cite{Giraud} (note that when we allow closed circuits, sometimes a~proper choice of the base tile could incidentally make the domain simply connected; see the $21_1$-examples in \cite[p.~5921]{OS}, and the corresponding domains~\cite{GT}). So we want the graph to be a tree. We refer to the upcoming paper \cite{KTI} for a more detailed discussion on that matter.
\end{itemize}
\end{Remark}

\subsection{Transplantability, operator groups and Schreier graphs}

\textbf{Transplantability.}
Two billiards constructed by tiling with respect to the matrices $M^{(\mu)}$, $N^{(\mu)}$
are said to be {\em transplantable} if there exists an invertible matrix~$T$~-- the {\em transplantation matrix}~-- such that
\[ TM^{(\mu)}  =  N^{(\mu)}T \qquad \mbox{for all}\ \ \mu.
\]

If the matrix $T$ is a permutation matrix, the two domains would just have the same shape.
One can show that transplantability implies isospectrality \cite{OS,GT}.

\begin{Remark}
All known counter examples are transplantable, see e.g.~\cite{GT}. (This follows from Remark \ref{rem}$(ii)$,$(iii)$ below.)
\end{Remark}

\textbf{Operator groups~-- group theoretical setting.}
Suppose $D$ is a (simply connected) Euclidean TI-domain on $N \in \mathbb{N}$ base $r$-gonal tiles, and  let $M^{(\mu)}$, $\mu \in \{1,2,\ldots,r\}$, be the corresponding permutation $N\times N$-matrices.  Def\/ine involutions $\theta^{(\mu)}$ on a set $X$ of $N$ letters $\Delta_1,\Delta_2,\ldots,\Delta_N$ (indexed by the base tiles) as follows:
$\theta^{(\mu)}(\Delta_i) = \Delta_j$ if $M^{(\mu)}_{ij} = 1$ and $i \ne j$. In the other cases, $\Delta_i$ is mapped onto
itself. Clearly, $\langle \theta^{(\mu)} \parallel \mu \in \{1,2,\ldots,r\}\rangle$ is a transitive permutation group
on $X$, which we call the {\em operator group} of $D$.

\begin{Remark}\label{rem}\qquad
\begin{itemize}\itemsep=0pt
\item[$(i)$]
Note that if two drums are transplantable, they have isomorphic operator groups ($T$ def\/ines the isomorphism in $\mathbf{GL}_N(\mathbb{C})$).
\item[$(ii)$]
In all
known examples, the operator groups are of a very restricted type:
they are all isomorphic to the classical group $\PSL_n(q)$, where $(n,q) \in \{(3,2),(3,3),(4,2),(3,4)\}$, cf.~\cite{Conw}.
\item[$(iii)$]
By combined work of O.~Giraud and K.~Thas, it follows that, conversely, all pairs arising from special linear group are the known ones.
This was observed in \cite{Giraud,KT,KT2,KT3}.
\end{itemize}
\end{Remark}

Let $D_1$ and $D_2$ be isospectral transplantable domains, on $N$ base $r$-gonal tiles.  Def\/ine the sets~$X_i$, indexed by the respective base tiles, as above.
Let $\{\theta^{(\mu)}\}$ be the aforementioned involutions def\/ined on $X_1$, and $\{\phi^{(\mu)}\}$ the involutions def\/ined on~$X_2$.  Then
\[  G_1 := \langle \theta^{(\mu)}\rangle \cong \langle \phi^{(\mu)}\rangle =: G_2.   \]

The action of $G_i$ on $X_i$, $i = 1,2$, can be identif\/ied with the left action $G_i \act G_i/H_i$ ($G_i/H_i$ is the left coset space of $H_i$ in $G_i$), where $H_i := {G_i}_{x_i}$ is
the stabilizer in $G_i$ of an arbitrary element~$x_i$ in~$X_i$.   The fact that $D_1$ and $D_2$ are transplantable translates in the fact that $G_1$ contains
a subgroup  $H' \cong H_2$ such that $H_1$ and $H'$  intersect every conjugacy class of $G_{1}$ in the same number of elements~\cite{GT}. We say they are {\em almost conjugate}. Note that conjugate subgroups always have this property.
For the sake of convenience, we denote $G_1$ by $G$ and $H_1$ by $H$.
We call a triple $(G,H,H')$ with the above properties a {\em Gassmann--Sunada triple}.

\textbf{The Schreier coset graph.}
The involution graph of $D_1$ (respectively $D_2$) which was earlier def\/ined is now nothing else than the so-called (undirected) ``Schreier coset graph'' of $G$ with respect to $H$ (respectively $H'$)
relative to $\{\theta^{(\mu)}\}$ (which is a generating set of involutions of~$G$), by def\/inition.
A {\em Schreier coset graph} is a Cayley graph where the vertices are cosets instead of elements.

\textbf{Permutation characters.}
Let $(G,H,H')$ be as above, with $H$ and $H'$
not conjugate. Then $G \act G/H$ and $G \act G/H'$ are inequivalent permutation representations of the same degree, and with the same permutation character (and vice versa,
 if $G/H$ and $G/H'$ are linearly equivalent $G$-sets, $(G,H,H')$ is a~Gassmann--Sunada triple).

\section{Classif\/ication of operator groups}

The question we now address (and which was f\/irst considered in \cite{KT3})  is:

\begin{Question}
Suppose $(D_1,D_2)$ is a transplantable pair of Euclidean TI-domains. Can one determine the associated
operator group, and by using the outcome, the pair $(D_1,D_2)$ itself?
\end{Question}

So the goal is to use the operator group in order  to determine the billiards. In the case of a $2$-transitive operator group we have Theorem~\ref{main}, the proof of which is the main objective of the present paper.

We use the following lemma throughout.

\begin{Lemma}[Y.~Okada and A.~Shudo~\cite{OS}]\label{Shudo}
All isospectral transplantable drums that unfold an $r$-gon $N$ times, are known if $N \leq 13$ $($and they all come from $\mathbf{PSL}$ groups$)$.
\end{Lemma}

We wish to obtain Theorem~\ref{main} largely as a corollary of a result on Schreier graphs of $2$-transitive groups.
More specif\/ically, we will show the following. (For the def\/initions of the groups considered below, we refer to~\cite{Atlas,DM}.)

\begin{Theorem}
\label{Schr}
Let $(G,X)$ be a $2$-transitive group, and $\Psi$ a generating set $($of $G)$ of involutions of size $r$ $(\geq 3)$. Suppose $G$ is not symmetric or alternating, not the Mathieu group $\mathbf{M}_{22}$, and also not a
symplectic group~$\mathbf{Sp}_{2m}(2)$.
The Schreier graph of $G$ with respect to $G_x$, $x \in X$ arbitrary, and relative to $\Psi$, cannot be a tree, unless $G \cong \PSL_n(q)$ with $(n,q) \in \{(3,2),(3,3),(4,2),(3,4)\}$.
\end{Theorem}

\begin{Corollary}[main result]\label{MR}
Let $D_1$ and $D_2$ be non-isometric isospectral simply connected domains in $\mathbb{R}^2$ which are transplantable.
If the associated operator group is $2$-transitive in its action on $N$ letters as defined by transplantability,
then it is isomorphic to $\PSL_n(q)$ with $(n,q) \in \{(3,2),(3,3),(4,2),(3,4)\}$, and
$(D_1,D_2)$ is one of the known pairs in~{\rm \cite{OS}}.
\end{Corollary}

\begin{proof}(We use the notation of the preceding sections.)
Consider the associated Gassmann--Sunada triple $(G,H,H')$ as earlier described. Let $\Psi = \{\theta^{(\mu)}\}$. Since our data must give rise to simply connected planar domains (essentially independently of the shape of the base tile),
the Schreier graph of~$G$ with respect to $H$ and relative to $\Psi$ is a connected tree, cf.\ Remark~\ref{remtree}$(ii)$.
Suppose $G$ is alternating or symmetric.

\textbf{The alternating groups $\bA_n$.}
In this case, the operator group $G$ is isomorphic to $\bA_n$, acting $2$-transitively and faithfully on
sets~$M$ and~$M'$
of~$m$ letters. Since we may suppose that~$m$ is suf\/f\/iciently large by Lemma~\ref{Shudo}, $m = n$, cf.~\cite{DM}.

Consider the two representations $(G,M)$ and $(G,M')$ and f\/ix $(x,y) \in M\times M'$; then~$G_x$ and~$G_{y}$ are almost conjugate  in~$G$, but not conjugate.
Obviously~\cite{DM}, we have that
\[ G_x \cong G_y \cong \bA_{n - 1},
\]
and both~$G_x$ and~$G_y$ are maximal subgroups of $G$ (as point-stabilizers in the respective
representations).

At this point it is convenient to recall the next theorem (in which $[U : V]$ denotes the index of $V$ in $U$, that is, $\vert U\vert/\vert V\vert$).

\begin{Theorem}[\protect{J.D.~Dixon and B.~Mortimer \cite[Theorem 5.2A]{DM}}]
Let $\bA = \bA_n$ be the alternating group on a set $\Omega$ of $n$ letters, $n \geq 5$, and let $r$ be an integer with
$1 \leq r \leq n/2$. Suppose that $G \leq \bA_n$ has index $[\bA_n : G] < \begin{pmatrix} n\\ r\end{pmatrix}$.
Then one of the following holds:
\begin{itemize}\itemsep=-1pt
\item[$(i)$]
For some $\Delta \subseteq \Omega$ with $\vert\Delta\vert < r$, we have $\bA_{[\Delta]} \leq G\leq \bA_{\{\Delta\}}$;
here, $\bA_{[\Delta]}$ is the pointwise stabilizer of $\Delta$ in $\bA$, while $\bA_{\{\Delta\}}$ is the setwise stabilizer.
\item[$(ii)$]
$n = 2m$ is even, $G$ is imprimitive with two blocks of size $m$, and $[\bA_n : G] = 1/2\begin{pmatrix} n\\ r\end{pmatrix}$.
\item[$(iii)$]
We are in six exceptional cases with $n \leq 9$~-- cf.~{\rm \cite{DM}}.
\end{itemize}
\end{Theorem}

It follows that $G_y$ also is a one-point-stabilizer in $(G,M)$, so that $G_x$ and $G_y$ are conjugate,
a contradiction.

By Lemma \ref{Shudo}, the cases in $(iii)$ and the groups $\mathbf{A}_n$ with $n \leq 4$ present no problem.

\textbf{The symmetric groups $\mathbf{S}_n$.}
The symmetric groups are handled similarly, and this follows from \cite[Theorem 5.2B]{DM}.

By Theorem~\ref{Schr}, we are left with the cases $G \cong \mathbf{M}_{22}$, and $G \cong \mathbf{Sp}_{2m}(2)$.
We noted in the previous sections that $G \act G/H$ and $G \act G/H'$ are inequivalent ($2$-transitive) permutation representations of the same degree, having the same character. This contradicts, for instance, \cite[Table, p.~8]{PJC}, when
$G$ is one of these groups\footnote{In fact, this is precisely the contradiction that we have obtained for the alternating and symmetric case. We could also have quoted the literature for that matter, but for the sake of completeness, we included the proof.}.
\end{proof}

\begin{proof}[Proof of Theorem~\ref{Schr}]
By the classif\/ication of f\/inite simple groups, the f\/inite 2-transitive groups are known. We brief\/ly give the list below.
A good description of them can be found in~\cite{DM,EG}. Moreover, also their $2$-transitive permutation representations are known, and this is a~fact that we need in our approach.

Let $\mathcal{G}$ be a f\/inite doubly transitive group.
Below, if $K$ is a group, $\mathrm{Aut}(K)$ denotes its automorphism group. Also, $q$ denotes any prime power, unless otherwise specif\/ied.
Then either (I)~$\mathcal{G}$~belongs to one of the following classes (the possible $2$-transitive permutation representations can be found in \cite{DM}; for def\/initions of the groups, see~\cite{Atlas}):
\begin{itemize}\itemsep=0pt
\item Symmetric groups $\mathbf{S}_{n}$, $n\geq 2$:
\item Alternating groups $\mathbf{A}_{n}$, $n\geq 4$;
\item Projective special linear groups $\PSL_{n}(q)\leq\mathcal{G}\leq \mathbf{P\Gamma L}_n(q)$, $n \geq 2$;
\item Symplectic groups $\mathbf{Sp}_{2m}(2)$, $m \geq 3$;
\item Projective special unitary groups $\mathcal{G}$, with $\mathbf{PSU}_3(q)\leq \mathcal{G} \leq \mathbf{P\Gamma U}_{3}(q)$;
\item Suzuki groups $\mathbf{Sz}(q)\leq \mathcal{G} \leq \mathrm{Aut}(\mathbf{Sz}(q))$ ($q = 2^{2e + 1}$, $e \geq 1$);
\item Ree groups $\mathbf{R}(q)\leq \mathcal{G} \leq \mathrm{Aut}(\mathbf{R}(q))$ ($q = 3^{2e + 1}$, $e \geq 0$);
\item Mathieu groups $\mathbf{M}_{11}$, $\mathbf{M}_{12}$, $\mathbf{M}_{23}$ and $\mathbf{M}_{24}$;
\item
the Mathieu group
$\mathbf{M}_{22} \leq \mathcal{G}\leq \mathrm{Aut}(\mathbf{M}_{22})$;
\item The Higman--Sims group $\mathbf{HS}$;
\item The Conway group $\mathbf{CO}_{3}$;
\end{itemize}
or
(II) $\mathcal{G}$ has a regular normal subgroup $\mathcal{N}$ which is elementary abelian of order $m=p^{d}$, where $p$ is a prime. In the latter case one can identify $\mathcal{G}$ with a group of af\/f\/ine transformations
\[ x\mapsto \sigma(x)+c\]
of $\mathbb{F}_{p^d}$. We call (II) the ``af\/f\/ine case''.

As already mentioned, the projective special linear groups were done in \cite{KT3}.
In a similar way one can start studying involutions in the modules on which the aforementioned groups act $2$-transitively, using similar equations, cf.~(\ref{Fixeq}) below. For most of the cases above this goes quite smoothly. Below we give a table which yields for each of the 2-transitive groups/representations~$\mathcal{G}$ the number $\phi(\mathcal{G})$, equal to the maximal number of points f\/ixed by an element which does not act as the identity, and the number of elements $N(\mathcal{G})$ contained in the module on which $\mG$ acts, see~\cite{EG}.

Suppose the group $\mG$ acts on a module with $N(\mG)$ elements and that we are unfolding $r$-gons, $r\geq 3$.
Then the general equation we have to solve is
\begin{gather}\label{Fixeq}
(r-2)N(G)=\sum_{j=1}^{r} \mbox{Fix}\big(\theta_{i}^{(j)}\big)-2,
\end{gather}
where Fix$(\theta^{(j)}_i)$ is the number of f\/ixed points of $\theta^{(j)}_i$, see~\cite{GT}. (Equation (\ref{Fixeq}) expresses the fact that the Schreier graph is a tree; it is exactly the same equation as that in \cite{Giraud}, but with the number of points of the space replaced by~$N(G)$.)
From this equation we get the following upper bound for $r$:
\begin{gather*}
r\leq \frac{2(N(\mG)-1)}{N(\mG)-\phi(\mG)}.
\end{gather*}

Therefore, we add the value $c(\mG)=\frac{2(N(\mG)-1)}{N(\mG)-\phi(\mG)}$. If $c(\mG)<3$ we get a contradiction, and this is indeed always the case. If $H$ is a group of the
f\/irst column, we have that $H \leq \mathcal{G}$, with $\mG$ as described in the list of the beginning of the proof.

\medskip

\centerline{\begin{tabular}{  c   c  c  c   c   }
\hline
Case & $H$ & $\phi(\mG)$ & $N(\mG)$ & $c(\mG)$ \tsep{2pt}\bsep{2pt}\\
\hline
1 &$\mathbf{PSU}_3(q)$ & $q+1$&$q^{3}+1$ & $\frac{2q^{3}}{q^{3}-q}$ \tsep{4pt} \\
2 & $\mathbf{Sz}(q)$ & $q+1$ & $q^{2}+1$&$\frac{2q^{2}}{q^{2}-q} $  \tsep{2pt}\\
3 & $\mathbf{R}(q)$ & $q+1$ & $q^{3}+1$& $\frac{2q^{3}}{q^{3}-q}$  \tsep{2pt}\\
4 & $\mathbf{M}_{11}$& 3 & 11& $\frac{5}{2}$\tsep{2pt}\\
5 & $\mathbf{M}_{11}$& 4 & 12& $\frac{11}{4}$\tsep{2pt}\\
6 & $\mathbf{M}_{12}$ & 4 & 12& $\frac{11}{4} $\tsep{2pt}\\
7 & $\mathbf{M}_{23}$ &7 & 23& $\frac{11}{4} $\tsep{2pt}\\
8 & $\mathbf{M}_{24}$ &8 & 24 &$\frac{23}{8}$\tsep{2pt}\\
9& $\mathbf{HS}$ & 16   & 176 &$\frac{35}{16}$ \tsep{2pt}\\
10 & $\mathbf{C0_{3}}$ & 36 & 276 &$\frac{55}{24}$ \tsep{2pt} \bsep{2pt}\\
\hline
\end{tabular}}

\medskip

Now we treat the  case which is left.

\textbf{The af\/f\/ine case.}
We discuss the dif\/ferent types of involutions that can occur in the automorphism group of a f\/inite
projective space, cf.~\cite{Segre}.
The reader can deduct the dif\/ferent types of involutions for af\/f\/ine spaces from this result.
\begin{itemize}\itemsep=0pt
\item
\textbf{Baer involutions.}
A {\em Baer involution} is an involution which is not contained in the linear automorphism group of the space, so that $q$ is a square, and it f\/ixes an
$n$-dimensional subspace over $\mathbb{F}_{\sqrt{q}}$ pointwise.
\item
\textbf{Linear involutions in even characteristic.}
If $q$ is even, and $\theta$ is an involution which is not of Baer type, $\theta$ must f\/ix an $m$-dimensional subspace of
$\PG(n,q)$ pointwise, with $1 \leq m \leq n \leq 2m + 1$.
In fact, to avoid trivialities, one assumes that $m \leq n - 1$.
\item
\textbf{Linear involutions in odd characteristic.}
If $\theta$ is a linear involution of $\PG(n,q)$, $q$~odd, the set of f\/ixed points is the union of
two disjoint complementary subspaces. Denote these by $\PG(k,q)$ and $\PG(n - k -1,q)$, $k \geq n - k - 1 > -1$.
\item
\textbf{Other involutions.}
All other involutions have no f\/ixed points.
\end{itemize}

We have to consider triples $(\mathbf{A},\{\theta^{(i)}\},r)$,
where $\mathbf{A}$ is a f\/inite af\/f\/ine space of dimension $n \geq 2$,
and $\{\theta^{(i)}\}$ a set of $r$ nontrivial involutory automorphisms of $\mathbf{A}$, satisfying
\begin{gather*}
 r(\vert \mathbf{A}\vert) - \sum_{j=1}^r\mbox{Fix}\big(\theta^{(j)}\big) = 2(\vert \mathbf{A}\vert - 1),
\end{gather*}
for the natural number $r \geq 3$. So we consider
\[ (r - 2)q^n + 2 = \sum_{j=1}^r\mbox{Fix}\big(\theta^{(j)}\big).         \]

Since an automorphic involution f\/ixes at most $q^{n - 1}$ points of $\mathbf{A}$, Lemma~\ref{Shudo}
 yields the desired contradiction.

The proof of Theorem \ref{Schr} is f\/inished.
\end{proof}

\section{Conclusion}

 The assumption of $2$-transitivity we made on the operator group $G$ allows us to use the  classif\/ication of f\/inite simple groups. However, it is a natural condition, since {\em all known counter examples are $2$-transitive groups}~-- they are even all $\PSL_n(q)$-groups. Moreover, for small va\-lues of $n$ it has been exhaustively checked by computer that there are no other examples. Not assuming any extra condition on $G$ yields the problem much harder and requires a more ge\-ne\-ral approach since the structural information is very limited in this case. Finding the right approach is the focus of ongoing work~\cite{KTI}. At present we suspect that operator groups {\em always} act $2$-transitively. If this is true a complete classif\/ication result for transplantable isospectral drums would be obtained by the main result of the present paper.

\subsection*{Acknowledgements}

The second author is partially supported by the Fund for Scientif\/ic Research~-- Flanders (Belgium).

\pdfbookmark[1]{References}{ref}
\LastPageEnding

\end{document}